\documentclass[aps,prd,reprint,showpacs,showkeys]{revtex4-1}
\usepackage{epsfig}
\usepackage{color}
\usepackage{hyperref}
\usepackage{url}

\def\NPB{{Nucl. Phys.} B}
\def\PLB{{Phys. Lett.}  B}
\def\PRL{Phys. Rev. Lett.}
\def\PRD{{Phys. Rev.} D}

\def\PRP{Phys. Rep.}

\def\JHEP{JHEP}

\def\JPG{J. Phys. G}
\def\IJMPA{ Int. J. Mod. Phys. A}

\def\CPC{Comput. Phys. Commun.}
\def\EPJC{Eur. Phys. J. C}
\def\RMP{Rev. Mod. Phys.}

\def\TMP{Theor. Math. Phys.}

\begin{document}

\title{Two-parametric model-independent observables for $Z'$ searching at the Tevatron}
\date{\today}

\author{A.V.~Gulov}
\email{alexey.gulov@gmail.com}
\affiliation{Dnipropetrovsk National University, Dnipropetrovsk, Ukraine}
\author{A.A.~Kozhushko}
\email{a.kozhushko@yandex.ru}
\affiliation{Dnipropetrovsk National University, Dnipropetrovsk, Ukraine}

\begin{abstract}
We propose a scheme of searches for the $Z'$ gauge boson as a virtual state in scattering processes at the Tevatron 
taking into account model-independent relations between the $Z'$ couplings to fermions. 
We integrate the Drell-Yan process cross setion to construct two-parametric observables, which are suitable for $Z'$ searches in the $p\bar{p} \to l^+ l^-$ process. The observables allow to constrain the $Z'$ vector and axial-vector couplings to SM fermions in a general parameterization with non-universal $Z'$ interactions with fermion generations. Also a one-parametric observable for searching for the popular leptophobic $Z'$ boson is proposed.
\end{abstract}


\maketitle

\section{\label{sec:intro}Introduction}

A new heavy neutral vector boson ($Z'$ boson) \cite{leike,*Lang08,*Rizzo06} is a popular scenario of searching for physics beyond the standard model (SM) of elementary particles in modern collider experiments.
Both the Tevatron and LHC collaborations try to catch the particle as a resonance in the Drell-Yan process. Observing no peak they conclude that the $Z'$ mass is no less than approximately 1.79 TeV \cite{CMS:2012zpr,*ATLAS:2011zpr} if one considers some predefined set of $Z'$ models. 

Another approach is to search for $Z'$ in processes where it manifests itself as a virtual state. This includes the processes with the so called low-energy neutral currents (LENC) mediated by a $Z$ boson. The contribution from a virtual $Z'$ state arises due to the $Z-Z'$ mixing. For example, the data on parity violation in cesium can be used to constrain the $Z'$ mass \cite{Derevianko2009}. A combined analysis of data on atomic parity violation, inelastic neutrino scattering, and neutrino-electron scattering \cite{Cho1998:1,*Cho1998:2,*Barger1998} allowed to constrain Fermi-like couplings that effectively represent $Z'$-mediated interactions at low energies \cite{Gulov:1999ry}. General review of low-energy constraints on the $Z'$ boson is presented in \cite{Erler:2009jh} and in Section 10 of Ref. \cite{rpp}.

Significant amount of the Tevatron data is collected at the $Z$-boson peak at 66-116 GeV. At these energies the $Z'$ boson also can manifest itself as an off-shell state, the $Z$ coupling constants are influenced by the $Z-Z'$ mixing, and these effects may allow to find $Z'$ signals by fitting the experimental data.

In order to select $Z'$ off-shell hints, proper observables have to be introduced to amplify possible signal \cite{Pankov98:1,*Pankov98:2,*Pankov98:3}. The signal generally means a deviation of some $Z'$ parameter (i.e., a coupling constant) from zero at a specified confidence level. The more parameters interfere in the observable, the weaker constraints on the parameters will be obtained. Thus, the key problem for off-shell $Z'$ detection is to maximally reduce the number of the $Z'$ couplings in the observable, which is used to fit the data. The ultimate scenario assumes a one-parametric observable. However, a two-parametric observable can be also useful and effective.
For example, the strategy to construct observables driven by one or two parameters was successfully applied to analyze the final data of the LEP experiment leading to model-independent hints and constraints on $Z'$ couplings \cite{Gulov:1999ry,Gulov:2005prd,*Gulov:2007prd}. So, attempts of selecting possible $Z'$ signals from Tevatron data seem to be perspective. 

In this paper we investigate possibilities of constructing few-parametric observables for the Drell-Yan process taking into account kinematics of the proton-antiproton collisions at $\sqrt{S} = 1.96$ TeV and model-independent para\-meterization of the $Z'$ couplings. Here we consider the case of a $Z'$ boson with non-universal $Z'$ couplings to fermion generations. The universality of couplings will be discussed in a separate paper, since it leads to a different (reduced) initial set of coupling constants and, consequently, requires a separate procedure of constructing few-parametric observables. We conclude that two-parametric observables exist at energies corresponding to $Z$ peak, and we obtain all of them. These observables can be used as a key to find possible signals of the off-shell $Z'$ boson. Data fitting is a subject of a separate investigation and lies beyond the scope of the paper.

The paper is organized as follows. In Section \ref{sec:abelianzpr} we provide all necessary information on the low-energy $Z'$ parameterization for our calculations. Section \ref{sec:ZprDY} contains specifics on $Z'$ contribution to the Drell-Yan process, uncertainties, and kinematic variables suitable for hadron colliders. In Section \ref{sec:theobservable} we construct the observables in a step-by-step manner. In Section \ref{sec:discussion} we briefly summarize and discuss the obtained results. The Appendix \ref{app:numerical} contains some supplemental numerical data.

% Z'

% RG Relations

\section{\label{sec:abelianzpr}Abelian $Z'$ couplings to leptons and quarks}

Being decoupled at energies of order of $m_Z$, the Abelian $Z'$ boson interacts with the SM particles as an additional $\tilde{U}(1)$ gauge boson. Its couplings to the SM fermions are usually parameterized by the effective Lagrangian:
\begin{eqnarray}\label{ZZplagr}
{\cal L}_{Z\bar{f}f}&=&\frac{1}{2} Z_\mu\bar{f}\gamma^\mu\left[
(v^\mathrm{SM}_{fZ}+\gamma^5 a^\mathrm{SM}_{fZ})\cos\theta_0 \right. \nonumber\\
&&\left. +(v_f+\gamma^5 a_f)\sin\theta_0 \right]f, \nonumber\\
{\cal L}_{Z'\bar{f}f}&=&\frac{1}{2} Z'_\mu\bar{f}\gamma^\mu\left[
(v_f+\gamma^5 a_f)\cos\theta_0 \right. \nonumber\\
&&\left. -(v^\mathrm{SM}_{fZ}+\gamma^5
a^\mathrm{SM}_{fZ})\sin\theta_0\right]f.
\end{eqnarray}
(Further details on the parameterization can be found in \citep{GulovSkalozub:2009review,*GulovSkalozub:2010ijmpa}.)
Here $f$ is an arbitrary SM fermion state;
$a_f$ and $v_f$ are the $Z'$ couplings to the
axial-vector and vector fermion currents, respectively; $\theta_0$
is the $Z$--$Z'$ mixing angle;
$v^\mathrm{SM}_{fZ}$,
$a^\mathrm{SM}_{fZ}$ are the SM couplings of the $Z$-boson. This
parameterization is suggested by a number of natural conditions:
\begin{itemize}
\item the $Z'$ interactions of renormalizable types are to
be dominant at low energies $\sim m_Z$. The non-renormalizable
interactions generated at high energies due to radiation
corrections are suppressed by the inverse heavy mass
$1/m_{Z^\prime}$ (or by other heavier scales $1/\Lambda_i\ll
1/m_{Z^\prime}$) and, therefore, at low energies can be neglected;
\item the $Z^\prime$ is the only neutral vector
boson with the mass $\sim m_{Z^\prime}$.
\end{itemize}

At low energies the $Z'$ couplings enter the cross section
together with the inverse $Z'$ mass, so it is convenient to
introduce the dimensionless couplings
\begin{equation}\label{avbar}
\bar{a}_f=\frac{m_Z}{\sqrt{4\pi}m_{Z'}}a_f,\quad
\bar{v}_f=\frac{m_Z}{\sqrt{4\pi}m_{Z'}}v_f,
\end{equation}
which are constrained by experiments.

Below the $Z'$ decoupling threshold the effective $\tilde{U}(1)$ symmetry is a trace of the renormalizability of an unknown complete model with the $Z'$ boson, and it leads to additional relations between the $Z'$ couplings 
% + epjc2000
\cite{GulovSkalozub:2009review,*GulovSkalozub:2010ijmpa}:
\begin{eqnarray}\label{RGrel1}
&&\bar{a}_{q_d} = \bar{a}_{l} = -\bar{a}_{q_u} = -\bar{a}_{\nu_l}=\bar{a},\nonumber\\
&&\bar{v}_{q_d} = \bar{v}_{q_u} + 2 \bar{a}, \qquad \bar{v}_{l} = \bar{v}_{\nu_l} + 2 \bar{a},
\end{eqnarray}
where $q_u$, $q_d$, $l$, and $\nu_l$  are an up-type and a down-type quark, a lepton, and a neutrino inside any fermion generation, correspondingly, and $\bar{a}$ is a universal constant, which defines also the $Z'$ coupling to the SM scalar fields and the $Z$--$Z'$ mixing angle in (\ref{ZZplagr}):
\begin{equation}\label{RGrel2}
\theta_0 \approx -2\bar{a}\frac{\sin \theta_W \cos
\theta_W}{\sqrt{\alpha_{\rm em}}} \frac{m_Z}{m_{Z'}}.
\end{equation}

As it was discussed in
\cite{GulovSkalozub:2009review,*GulovSkalozub:2010ijmpa}, the
relations (\ref{RGrel1}) cover a popular class of models based on
the ${\rm E}_6$ group (the so called LR, $\chi$-$\psi$ models). Thus, they
describe correlations between $Z'$ couplings for a wide set of
models beyond the SM. That is the reason to call the relations
model-independent ones.

As a result, $Z'$ couplings can be parameterized by seven
independent constants $\bar{a}$, $\bar{v}_u$, $\bar{v}_c$,
$\bar{v}_t$, $\bar{v}_e$, $\bar{v}_\mu$, $\bar{v}_\tau$.
These parameters must be fitted in
experiments. In a particular model, one has some specific values
for them. In case when the model is unknown, these parameters
remain potentially arbitrary numbers. 

% Drell-Yan
\section{\label{sec:ZprDY}Abelian $Z'$ in the Drell-Yan process}

\begin{figure}[t]
\centering{
\includegraphics[width=0.9\linewidth]{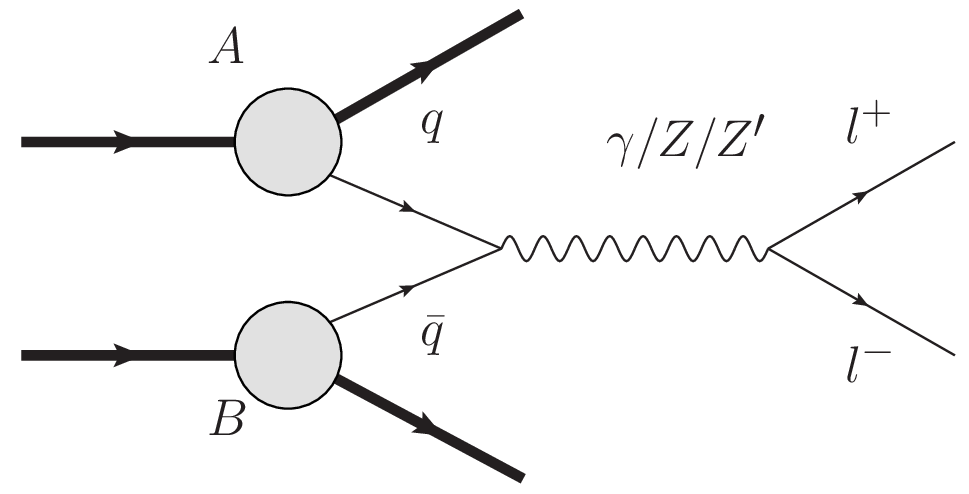}}
\caption{\label{fig:DY}Drell-Yan scattering process. $A$ and $B$ mark the interacting hadrons -- $p$ and $\bar{p}$ in the Tevatron case.}
\end{figure}

At the Tevatron the most prominent signal of the Abelian $Z'$ boson is expected in the $p\bar{p} \to l^+ l^-$ scattering process (Fig. \ref{fig:DY}). The general idea of our approach is equally applicable both for dielectrons and dimuons in the final state. To be definite, we shall consider the dielectron case. 
%The dimuon pseudorapidity range covered by a detector is different from the dielectron case. 
Specifics concerning the dimuon final state will be addressed to in Section \ref{sec:theobservable} and in the Appendix\ref{app:numerical}.
The cross section of this process can be written in form of the partonic cross sections combined with the parton distribution functions (PDFs):
\begin{eqnarray}
\frac{\partial^3 \sigma_{AB}}{\partial x_q \partial x_{\bar{q}} 
\partial \hat{t}} &=&
\sum_{q,\bar{q}}\, f_{q,A}
(x_q,Q^2)f_{\bar{q},B}(x_{\bar{q}},Q^2)
\frac{\partial \sigma_{q\bar{q}\to e^+ e^-}}{\partial \hat{t}},\nonumber\\
\sigma_{q\bar{q}\to e^+ e^-} &=& \sigma_{q\bar{q}\to e^+ e^-}(\hat{t}),
\end{eqnarray}
where $A$, $B$ mark the interacting hadrons ($p$ or $\bar{p}$)
with the four-momenta $k_A$, $k_B$; $f_{q,A}(x_q, Q^2)$ is the PDF for the parton $q$ in the hadron $A$ with
the momentum fraction $x_q$ ($0 \leq x_q \leq 1$) at the factorization
scale $Q^2$. To access the parton distribution data, we use the MSTW 2008 package \cite{mstw:1,*mstw:2}. 
The quantity $\sigma_{q\bar{q}\to e^+ e^-}$ is the parton-level 
cross section, which depends on the Mandelstam variable $\hat{t} = 
(p_{e^+} - p_{q})^2$. All parton-level calculations are performed using \textit{FeynArts} \cite{FeynArts:1,*FeynArts:2} and \textit{FormCalc} \cite{FormCalc:1,*FormCalc:2} packages. Hereafter, the hat over a variable denotes that this variable refers to the parton-level cross section. 

We define the PDF factor for each quark flavor:
\begin{eqnarray}
f_{q,A}(x_q,Q^2)f_{\bar{q},B}(x_{\bar{q}},Q^2) = F_{q\bar{q}}(x_q,x_{\bar{q}},Q).
\end{eqnarray}
The PDF factor and the parton-level cross section are calculated
in the leading order (LO) in $\alpha_S$, and $\sigma_{AB}$ in the LO is obtained in this way. The next-to-next-to-leading
order (NNLO) corrections are then taken into account by multiplying
$\sigma_{AB}$ by the NNLO K-factor, which is calculated using the \textit{Vrap} software \cite{Dixon:vrap} (see also ref. \cite{Dixon:2004prd}).

We also consider two kinds of uncertainties: 
\begin{itemize}
\item the PDF uncertainties $\Delta\sigma_{\mathrm{PDF}}$. The MSTW 2008 package provides
68\% CL and 90\% CL intervals. We consider the latter one;
\item the uncertainties due to the factorization scale variation, $\Delta\sigma_{Q}$. To incorporate these uncertainties, we follow the common procedure and vary $Q$ from $\sqrt{\hat{s}}/2$ to $2\sqrt{\hat{s}}$, where $\hat{s}$ is a Mandelstam variable for the partonic level process: $\hat{s} = (p_{e^+} + p_{e^-})^2$.
\end{itemize} 

The cross section then can be written as $\sigma_{\mathrm{DY}} \pm~\Delta\sigma_{\mathrm{PDF}}\pm\Delta\sigma_{Q}$.

The obtained triple-differential cross section provides full description for the Drell-Yan process. It is expressed in terms of three kinematic variables: $x_q$, $x_{\bar{q}}$, and $\hat{t}$. The shortcoming of these variables is that all three of them enter both the PDF multiplier and the parton-level cross section, since $\hat{s}$ is not an independent value ($\hat{s} = x_q x_{\bar{q}} S$).

The quantities that are directly measured in experiments and used for event selection are the pseudorapidities $\eta_{\pm}$ and transverse momenta $p_{T}^{\pm}$ of the final-state electrons. In the leading order in $\alpha_S$ the relation $p_{T}^+ = -p_{T}^- = p_T$ applies.
The Mandelstam variables $\hat{s}$, $\hat{t}$ and the momentum fractions $x_q$, $x_{\bar{q}}$ are expressed as
\begin{eqnarray}
&& \hat{s} = M^2 = 4 p_T^2 \, \cosh^2 \, \frac{\eta_{+}-\eta_{-}}{2}, \quad
\hat{t} = -\frac{M^2}{1+ e^{(\eta_{+}-\eta_{-})/2}}, \nonumber\\
&& x_q = \frac{M}{\sqrt{S}} e^{(\eta_{+}+\eta_{-})/2}, \quad
x_{\bar{q}} = \frac{M}{\sqrt{S}} e^{-(\eta_{+}+\eta_{-})/2}.
\end{eqnarray}
Note, that $x_q$ and $x_{\bar{q}}$ depend only on the sum of the electron pseudorapidities, $Y = (\eta_{+}+\eta_{-})/2$, while $\hat{t}$ is expressed in terms of the difference of the pseudorapidities, $y = (\eta_{+}-\eta_{-})/2$. The $Y$ variable is the well-known intermediate-state rapidity, while $y$ is related to the scattering angle in the $q\bar{q} \to e^+ e^-$ process and governs the parton-level kinematics (it can be found introduced in some textbooks, for example in \cite{PeskinSchroeder}). In this way the cross section is obtained as a function of $M$, $Y$, $y$:
\begin{eqnarray}
\label{eq:MYy_cs}
\frac{\partial^3 \sigma_{AB}}{\partial M \partial Y 
\partial y} &=&
\sum_{q,\bar{q}}\, F_{q\bar{q}}(M,Q^2,Y)
\frac{\partial \sigma_{q\bar{q}\to e^+ e^-}}{\partial y},\nonumber\\
\sigma_{q\bar{q}\to e^+ e^-} &=& \sigma_{q\bar{q}\to e^+ e^-}(M,y).
\end{eqnarray}
Here, the $Q$-dependence of $F_{q\bar{q}}$ is shown just to indicate that we incorporate the uncertainties due to the scale variation in our analysis.

Leading $Z'$ contribution to the Drell-Yan process arises from interference between diagrams with $\gamma^*/Z$ and $Z'$ intermediate states, resulting in corrections of order of $O(\tilde{g}^2)$. The cross section reads as
\begin{eqnarray}
\label{eq:zpr_factors}
\sigma_\mathrm{DY} &=& \sigma_\mathrm{SM} + \sigma_{Z'}, \nonumber\\
\sigma_{Z'} &=& \bar{a}^2 \sigma_{\bar{a}^2} + \bar{a} \bar{v}_e \sigma_{\bar{a} \bar{v}_e} 
+ \bar{a} \bar{v}_u \sigma_{\bar{a} \bar{v}_u} + 
\bar{v}_u \bar{v}_e \sigma_{\bar{v}_u \bar{v}_e} \nonumber\\
&& + \bar{a} \bar{v}_c \sigma_{\bar{a} \bar{v}_c} + 
\bar{v}_c \bar{v}_e \sigma_{\bar{v}_c \bar{v}_e}.
\end{eqnarray}
Here $\bar{a}$, $\bar{v}_f$ are the couplings defined in (\ref{avbar}), (\ref{RGrel1}), and $\sigma_{\bar{a}^2}$, $\sigma_{\bar{a} \bar{v}_f}$, $\sigma_{\bar{v}_f \bar{v}_{f'}}$ are the numerical factors that depend on $M$, $Y$, $y$. In this approximation there are six independent unknown quantities entering the Drell-Yan process cross section.
In (\ref{eq:zpr_factors}) the factors that include $\bar{v}_u$ and $\bar{v}_c$ arise only due to contributions of first and second generation fermions, respectively. The contribution from the third generation is neglected due to the nature of (anti)protons.

Once again, we note that $Y$ enters the PDF factors only, while $y$ is included into the parton-level cross sections only. This is a crucial point for our analysis, as it allows us to treat $F_{q\bar{q}}$ and $\sigma_{q\bar{q}\to e^+ e^-}$ separately.
Therefore, we can try to use any peculiarities in the $M$-, $Y$-, and $y$-dependence of the PDF factors and partonic cross sections to suppress some of the numerical factors in (\ref{eq:zpr_factors}). For example, in case after integration by one of the kinematic variables over some specific region the factor $\sigma_{\bar{v}_c \bar{v}_e}$ appears to be much smaller than the other factors, we may neglect its contribution to the cross section and deal with five unknown parameters instead of six we had initially. Of course, we assume that all the combinations of the $Z'$ couplings in the cross section are of the same order of magnitude. The leptophobic $Z'$ case, which seems to be a very popular parameterization nowadays, is treated separately in Sec. \ref{sec:theobservable}.

In addition to the $Z'$ couplings, there are another two unknown $Z'$ parameters that affect $\sigma_\mathrm{DY}$. These are the $Z'$ mass $m_{Z'}$ and decay width $\Gamma_{Z'}$. The latest data from the CMS and ATLAS indicates that $Z'$ is heavier than 1.79 TeV. This means, that for energies close to the $Z$ peak the $\sigma_\mathrm{DY}$ dependencies on $m_{Z'}$ and $\Gamma_{Z'}$ can be neglected, assuming that the $Z'$ peak is far away. 

The $Y$ and $y$ values that we can investigate are limited by detector performance and conservation laws.
%The regions of values of the two kinematic variables, $Y$ and $y$, are limited by detector performance and conservation laws. 
From the condition $0 \leq x_{q,\bar{q}} \leq 1$ it is easy to obtain the $M$-dependent limits
\begin{eqnarray}
\label{eq:conslim}
-\ln \frac{\sqrt{S}}{M} \leq Y \leq \ln \frac{\sqrt{S}}{M}.
\end{eqnarray}

The electromagnetic calorimeters of both Tevatron detectors, CDF and D0, cover the electron pseudorapidity range $|\eta_\pm| \leq 3.2$ \cite{CDF_DO_rapidity_ee}.
%\begin{eqnarray}
%\label{eq:CDFlimits}
%|\eta| \leq 3.2. \nonumber
%\end{eqnarray}
Therefore,
\begin{eqnarray}
\label{eq:detlim}
|Y|\leq 3.2.
\end{eqnarray}
The limits for $y$ are the same as for $Y$.

This section can be briefly summarized by saying the following: the cross section of the Drell-Yan process contains six unknown linear-independent terms inspired by $Z'$ boson. The cross section depends on three kinematic variables, which will be used in what follows to suppress some of the contributions from the unknown $Z'$ parameters. This will allow us to amplify the signal of $Z'$ that is possibly hidden in the Tevatron experimental data.

\section{\label{sec:theobservable}The Observable}

Of course, the most detailed description of a scattering process is contained in the differential cross section. But a possible $Z'$ signal can be washed out by the interference between the six independent combinations of $Z'$ couplings entering the cross section. In general, integration by kinematic variables can leave this situation without changes. We need to pay special attention to the integration scheme to reduce the number of interfering parameters in order to make a successful data fit possible. This scheme must derive benefits from kinematic properties of the cross section.

\subsection{\label{subsec:PDF}Integrating by $Y$}

The intermediate-state rapidity $Y$ enters the PDF factors only. Let us study the $M$- and $Y$-dependence of $F_{q\bar{q}} (M,Y)$ in Eq. (\ref{eq:MYy_cs}). At any fixed kinematically allowed $Y$ value $F_{q\bar{q}}$ is a smooth monotonically decreasing function of $M$.
Kinematic properties of $F_{q\bar{q}}$ are different for each flavor but independent of $Z'$ properties. 
So, the $Y$-dependence of the cross section can be utilized to suppress the contributions of the second generation, i.e., the terms with $\bar{a}\bar{v}_c$ and $\bar{v}_c \bar{v}_e$ in Eq. (\ref{eq:zpr_factors}). 

\begin{figure}[t]
\centering{\includegraphics[width=1\linewidth]{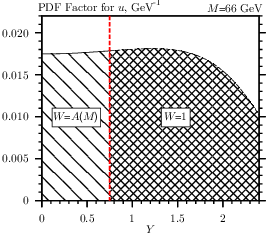}}
\caption{\label{fig:IntegrationDemo}Plot illustrating the integration scheme with the weight function from Eq. (\ref{eq:observ_Y_wf}). The vertical dashed line represents $Y_1$. In this particular case $Y_1$ is set to 0.75, and the upper integration limit $Y_m$ is 2.35.}
\end{figure}

We use the following integration scheme
\begin{eqnarray}
\label{eq:observ_Y}
\sigma_1 = \int_{-Y_m}^{Y_m} dY \, W(M,Y)\frac{\partial^3 \sigma_{\mathrm{DY}}}{\partial Y \partial M \partial y}
\end{eqnarray}
 with a simple piecewise-constant weight function
\begin{eqnarray}
\label{eq:observ_Y_wf}
W(M,Y) = \left\{\begin{array}{ll}
A(M),& 0<|Y|\le Y_1,\\ 
1,& Y_1<|Y|<Y_m.\end{array}\right.
\end{eqnarray}
In Eq. (\ref{eq:observ_Y}) $\sigma_1$ denotes the value obtained by the integration of the triple-differential cross section $\sigma_{\mathrm{DY}}$ by $Y$. 
In fact, we integrate the PDF factor in Eq. (\ref{eq:MYy_cs}):
\begin{eqnarray}
\label{eq:PDF_int_Y}
F_{q\bar{q}} (M) &=& 2 \int_0^{Y_m} \,  dM \, W(M,Y) \,\nonumber\\
 &&\times F_{q\bar{q}} (M,Y),\\
\sigma_1 &=& \sum_{q,\bar{q}} F_{q\bar{q}}(M)\frac{\partial^2 \sigma_{q\bar{q}\to e^+ e^-}}{\partial M \partial y}.
\end{eqnarray}
So, in in this subsection we will study $F_{q\bar{q}} (M)$ for different quark generations.
An example of this integration scheme is illustrated in Fig. \ref{fig:IntegrationDemo}.
Note, that the $Y$-distribution for the Drell-Yan cross section is symmetric. 

The $Y_m$ value is some positive boundary chosen for the $Y$ integration region.
The exact maximal $Y_m$ could be determined from Eq. (\ref{eq:conslim}). However, this would be technically inconvenient. First, it depends on $M$. Second, it leads to difficulties in usage of the K-factor. We calculate the K-factor as
\begin{eqnarray}
K(M,Y) = \frac{\sigma_\mathrm{SM}^\mathrm{NNLO}}{\sigma_\mathrm{SM}^\mathrm{LO}}.
\end{eqnarray}
For different considered regions of $M$ values and for $Y$ close to boundary values (\ref{eq:conslim}) the numerical uncertainty of the calculations becomes large for $\sigma_\mathrm{SM}^\mathrm{LO}$. Because of this the K-factor becomes a non-monotonic fast-varying function of $Y$ and, therefore, cannot be used to improve the new physics contributions to the cross section.
To avoid these difficulties, we chose a somewhat lower value of $Y_m$ being also independent on $M$.

\begin{figure*}[t]
\centering{
\includegraphics[width=0.49\linewidth]{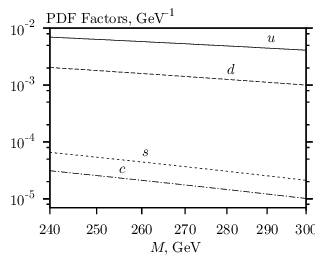}
\includegraphics[width=0.49\linewidth]{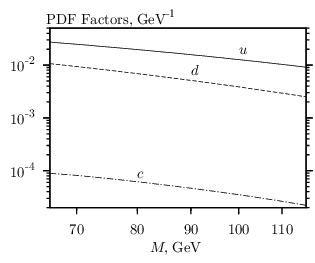}\\
\hfill (a) \hfill\hfill (b) \hfill}
\caption{\label{fig:ZpeakIntegrated}Plots illustrating the suppression of the contributions of second-generation quarks to the Drell-Yan cross section in different regions of $M$ values and with different integration schemes. The plotted values are $F_{u\bar{u}}$, $F_{d\bar{d}}$, $F_{c\bar{c}}$, and $F_{s\bar{s}}$ (not shown on the right plot. Because of the utilized integration scheme at some $M$ is becomes negative, but its absolute value is even smaller than $F_{c\bar{c}}$) integrated by $Y$: a) The integration by $Y$ is carried out over the region $|Y| \leq 1.65$ with $A(M)=1$; b) the integration by $Y$ is carried out over the region $|Y| \leq 2.35$ with $A(M)$ from Fig. \ref{fig:AvsM}, where $Y_1$ is set to 0.75.}
\end{figure*}

First, let us consider the unit weight function (i.e., $A(M)=1$). In this case there is no effect from $Y_1$. The integration limit $Y_m$ is set to $1.65$, which corresponds to a bin bound at the D0 detector (see, for example, \cite{D0_zpr_y}). The plots of $F_{q\bar{q}} (M)$ for $u$, $d$, $c$, and $s$ quarks are presented in Fig. \ref{fig:ZpeakIntegrated} (a). These plots indicate that for $M > 240 \mathrm{~GeV}$ the factors for the second-generation quarks amount to less than 1\% of those for the first generation.
This leads to the conclusion that for these values of $M$ we can use the standard integration with $A(M)=1$ neglecting the contributions from the second generation, $\sigma_{\bar{a} \bar{v}_c}$ and $\sigma_{\bar{v}_c \bar{v}_{l}}$.

Now consider the $M$ values at the $Z$-peak. Both CDF and D0 collaborations define limits of this region to be symmetric with respect to the $Z$ boson mass. These limits are often set to either $66 \mathrm{~GeV} \leq M \leq 116 \mathrm{~GeV}$ or $71 \mathrm{~GeV} \leq M \leq 111 \mathrm{~GeV}$ \cite{CDF66,D071}. In the present paper the former alternative is used. Actually, the choice of specific lower and upper limits does not affect our results. There are only two general requirements: the limits have to be symmetric with respect to $m_Z$ and large enough so that we could set all quark masses to zero.

\begin{figure*}[t]
\centering{
\includegraphics[width=0.32\linewidth]{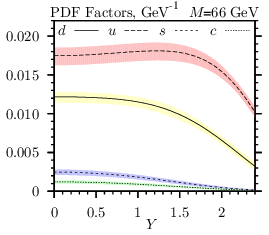}
\includegraphics[width=0.32\linewidth]{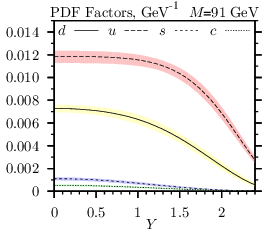}
\includegraphics[width=0.32\linewidth]{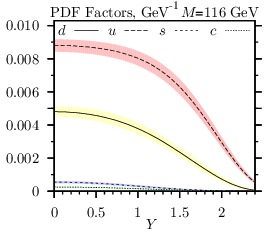}}
\caption{\label{fig:PDFsY}Plots for $F_{q\bar{q}} (M,Y)$ versus $Y$ at different $M$ values. The uncertainties that arise from the PDF errors and factorization scale variation are also shown (see Sec. \ref{sec:ZprDY}).}
\end{figure*}

In Fig. \ref{fig:PDFsY} the plots for $F_{q\bar{q}} (M,Y)$ versus $Y$ at different $M$ values are shown for $u$, $d$, $c$, and $s$ quarks. The relative contributions of second generation quarks amount up to 11\% at $M = 66 \mathrm{~GeV}$ and cannot be neglected. There is a qualitative difference between the PDF factors for the first and second generations. At some energies the factor for $u$ quarks is convex for $Y$ close to zero (at somewhat lower energies this is also the case for $d$). This is due to the nature of a proton.

\begin{figure}[t]
\centering{
\includegraphics[width=1\linewidth]{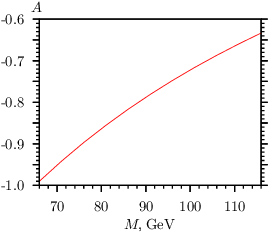}}
\caption{\label{fig:AvsM}Weight coefficient $A(M)$ used for integration over the $Z$-peak region. $Y_m=2.35$, $Y_1=0.75$.}
\end{figure}

For any given $M$ value from the $Z$-peak region we can adjust the weight function in such a way that the factors $F_{c\bar{c}} (M)$ and $F_{s\bar{s}} (M)$ amount to less than 1\% of each of the factors $F_{u\bar{u}} (M)$ and $F_{d\bar{d}} (M)$:
\begin{eqnarray}
F_{c\bar{c},\,s\bar{s}} (M) \leq 0.01 F_{u\bar{u},\,d\bar{d}} (M)
\end{eqnarray}
This is shown on Fig. \ref{fig:ZpeakIntegrated} (b).
Therefore, the contributions of the second generation-fermions are suppressed, and again $\sigma_{\bar{a} \bar{v}_c}$ and $\sigma_{\bar{v}_c \bar{v}_{e}}$ are excluded from $\sigma_\mathrm{DY}$. The weight coefficient $A(M)$ can be determined for several $M$ values and interpolated in the $Z$-peak region (see the Appendix\ref{app:numerical}). For our specific case $A(M)$ is plotted in Fig.~\ref{fig:AvsM}. Here $Y_m$ is set to $2.35$, and $Y_1$ is $0.75$.

As a result, we obtain the cross section $\sigma_1$, which depends on $y$ and $M$ and contains four linearly independent $Z'$ terms instead of six:
\begin{eqnarray}
\label{eq:sigma1}
\sigma_1 &=& \sigma_\mathrm{1\,SM} + \bar{a}^2 \sigma_{1\,\bar{a}^2} + \bar{a} \bar{v}_e \sigma_{1\,\bar{a} \bar{v}_e} \nonumber\\
&& + \bar{a} \bar{v}_u \sigma_{1\,\bar{a} \bar{v}_u} + 
\bar{v}_u \bar{v}_e \sigma_{1\,\bar{v}_u \bar{v}_e}.
\end{eqnarray}

Our next step is to use the remaining two kinematic variables, $M$ and $y$, to get rid of another two unknown combinations of the $Z'$ couplings.

\subsection{\label{subsec:partonic}Integrating by $M$ and $y$}

The difference of the pseudorapidities, $y$, enters the parton-level cross section of the Drell-Yan process, $\sigma_{q\bar{q}\to e^+ e^-}$, only and is irrelevant for the PDF analysis. The parton-level cross section depends also on $M$ through four `resonant' functions:
\begin{widetext}
\begin{eqnarray}
\label{eq:RF}
%f_1(M) &=& \frac{1}{(M^2/m_Z^2 - 1)^2 + \Gamma_Z^2 /m_Z^2}, \nonumber\\
%f_2(M) &=& \frac{(M^2/m_Z^2 - 1)}{(M^2/m_Z^2 - 1)^2 + \Gamma_Z^2 /m_Z^2},
%\nonumber\\
%f_2'(M) &=& \frac{(M^2/m_{Z'}^2 - 1)}{(M^2/m_{Z'}^2 - 1)^2 +
%\Gamma_{Z'}^2/m_{Z'}^2},
%\nonumber\\
%f_3(M) &=& \frac{\frac{M^2 \Gamma_Z \Gamma_{Z'}}{m_Z^3 m_{Z'}}+
%\frac{M^2}{m_Z^2}(\frac{M^2}{m_Z^2}-1)(\frac{M^2}{m_{Z'}^2}-1)}
%{((M^2/m_Z^2 - 1)^2 + \Gamma_Z^2 /m_Z^2)((M^2/m_{Z'}^2 - 1)^2 +
%\Gamma_{Z'}^2/m_{Z'}^2)} .
f_1(M) = \frac{1}{(M^2/m_Z^2 - 1)^2 + \Gamma_Z^2 /m_Z^2}, \quad
f_2(M) = \frac{(M^2/m_Z^2 - 1)}{(M^2/m_Z^2 - 1)^2 + \Gamma_Z^2 /m_Z^2}, \quad
f_2'(M) &=& \frac{(M^2/m_{Z'}^2 - 1)}{(M^2/m_{Z'}^2 - 1)^2 +
\Gamma_{Z'}^2/m_{Z'}^2},
\nonumber\\
f_3(M) = \frac{M^2 \Gamma_Z \Gamma_{Z'}/(m_Z^3 m_{Z'})+
(M^2/m_Z^2)(M^2/m_Z^2-1)(M^2/m_{Z'}^2-1)}
{[(M^2/m_Z^2 - 1)^2 + \Gamma_Z^2 /m_Z^2][(M^2/m_{Z'}^2 - 1)^2 +
\Gamma_{Z'}^2/m_{Z'}^2]} .
\end{eqnarray}
\end{widetext}
Here $m_{Z, Z'}$ and $\Gamma_{Z, Z'}$ denote the masses and the widths of the $Z$ and $Z'$ bosons. We investigate the energy region close to the $Z$ boson peak. As it was noted earlier, in this case we do not care about the specific values of the $Z'$ mass and decay widths. But at this point for numerical calculations we are going to set specific values for $m_{Z'}$ and $\Gamma_{Z'}$. Following the recent LHC results \cite{CMS:2012zpr,*ATLAS:2011zpr}, we set $m_{Z'}$ to 1.8 TeV and assume the decay width to be 10\% of the mass. That is, we use some asymptotics of $f'_2$ and $f_3$ at $M\ll m_{Z'}$.

\begin{figure}[t]
\centering{
\includegraphics[width=1\linewidth]{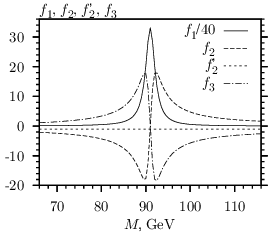}}
\caption{\label{fig:RF}Plots for the resonant functions, which are given by Eqs. \ref{eq:RF}, in the region $66 \mathrm{~GeV} \leq M \leq 116 \mathrm{~GeV}$.}
\end{figure}

As it can be seen from Fig. \ref{fig:RF}, the $f_1$ function is dominant. 
The functions $f_2$, $f'_2$ are odd-like with respect to $M=m_Z$, and the function $f_3$ is small.
As a consequence, after integrating by $M$ over the discussed symmetric $Z$-peak region the functions $f_2$, $f'_2$, and $f_3$ are negligible compared to $f_1$. 
We are going to use the discussed feature in what follows.

\begin{figure*}[t]
\centering{
\includegraphics[width=0.49\linewidth]{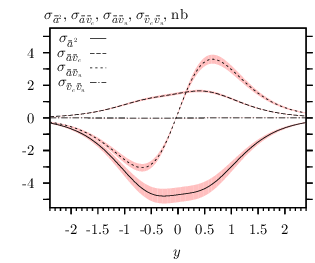}
\includegraphics[width=0.49\linewidth]{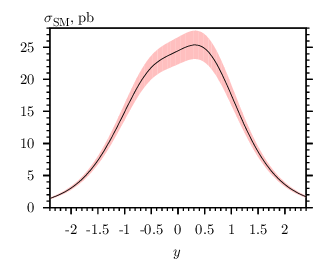}\\
\hfill (a) \hfill\hfill (b) \hfill}
\caption{\label{fig:factors}Plots for (a) the $Z'$-related factors and (b) $\sigma_\mathrm{2\,SM}$ from Eq. (\ref{eq:sigma2}). The uncertainty bands are also shown.}
\end{figure*}

When investigating the $M$-dependence of the hadronic cross section $\sigma_1$, we deal not with the resonant functions themselves, but with their products with the PDF factors.
The general form of $\sigma_1$ can be written as
\begin{eqnarray}
\label{eq:sigma1_yM}
\sigma_1 - \sigma_\mathrm{1\,SM}= \frac{\cosh 2y}{\cosh^4 y} \left[ a(M) \tanh 2y + b(M) \right],
\end{eqnarray}
where $a(M)$ and $b(M)$ are some functions that include the unknown couplings $\bar{a}$, $\bar{v}_u$, and $\bar{v}_e$.
%[see Eq. (\ref{eq:sigma1})]
The $M$-dependence arises from the `resonant' functions multiplied by $F_{q\bar{q}}(M)$ from Eq. (\ref{eq:PDF_int_Y}). From the plots in Fig.~\ref{fig:ZpeakIntegrated} we can conclude that the factors $F_{q\bar{q}}(M)$ are smooth, monotonic, and slowly-varying in the considered region. 
Therefore, we stress that all the discussed properties of $f_1$, $f_2$, $f'_2$, and $f_3$ are generally maintained, when these functions are multiplied by $F_{q\bar{q}}(M)$.

Naturally, $f'_2$ and $f_3$ do not enter the SM part $\sigma_\mathrm{1\,SM}$. There are four factors entering the $Z'$ contribution: $\sigma_{1\,\bar{a}^2}$, $\sigma_{1\,\bar{a} \bar{v}_e}$, $\sigma_{1\,\bar{a} \bar{v}_u}$, and $\sigma_{1\,\bar{v}_u \bar{v}_e}$ [see Eq. (\ref{eq:sigma1})]. The factor $\sigma_{1\,\bar{v}_u \bar{v}_e}$ does not depend on $f_1$, and, therefore, according to our discussion of properties of the `resonant' functions we may eliminate it by the straightforward integration by $M$ over the $Z$-peak region (66 GeV $\leq \, M \, \leq$ 116 GeV). The resulting value is denoted $\sigma_2$:
\begin{eqnarray}
\label{eq:sigma2}
\sigma_2 - \sigma_\mathrm{2\,SM} & = & \int dM \, (\sigma_1 - \sigma_\mathrm{1\,SM}) \nonumber\\ 
&=& \frac{\cosh 2y}{\cosh^4 y} \left( a \tanh 2y  + b \right), \nonumber\\
\sigma_2 & = & \sigma_\mathrm{2\,SM} + \bar{a}^2 \sigma_{2\,\bar{a}^2} + \bar{a} \bar{v}_e \sigma_{2\,\bar{a} \bar{v}_e} + \bar{a} \bar{v}_u \sigma_{2\,\bar{a} \bar{v}_u},
\nonumber\\
a &=& \int dM \, a(M), \qquad b = \int dM \, b(M).
\end{eqnarray}
The factors $\sigma_\mathrm{2\,SM}$, $\sigma_{2\,\bar{a}^2}$, $\sigma_{2\,\bar{a} \bar{v}_e}$, $\sigma_{2\,\bar{a} \bar{v}_u}$, and $\sigma_{2\,\bar{v}_u \bar{v}_e}$ are plotted on Fig. \ref{fig:factors}. It can be seen that $\sigma_{2\,\bar{v}_u \bar{v}_e}$ is negligibly small compared to the other three factors indicating that our assumption is relevant.

We are not concerned about $\sigma_\mathrm{2\,SM}$ at the moment and shall turn to investigating the $y$-dependence of the $Z'$-related contribution presented in Eq.~(\ref{eq:sigma2}).
The behavior of the $\sigma_{\bar{a} \bar{v}_u}$ factor is governed by its odd part, while the $\sigma_{\bar{a}^2}$ and $\sigma_{\bar{a} \bar{v}_e}$ factors are obviously dominated by their even parts.
From the plots on Fig. \ref{fig:factors} (a), one can conclude that it is possible to suppress one of the three factors by integrating the cross section by $y$ over a symmetric region. Remember, that the integration limits for $y$ are the same as for $Y$. In our case 
\begin{equation}
\label{eq:yregion}
-2.35 \leq y \leq 2.35.
\end{equation}
For example, we can integrate them with a piecewise-constant function
\begin{eqnarray}
\label{eq:example_weight}
\omega(y) = \left\{\begin{array}{ll}
x,& y \geq 0,\\ 
1,& y<0.\end{array}\right.
\end{eqnarray}
Here $x$ is some real number. The sign of $x$ is chosen depending on which specific factor we want to suppress. The resulting observable $\sigma^*$ would be a somewhat modified forward-backward scattering asymmetry:
\begin{eqnarray}
\sigma^* = \int dy \, \omega(y) \, \sigma_2.
\end{eqnarray}

We propose an approach that is a bit more general. The weight function that we use has a structure similar to the one in Eq. (\ref{eq:sigma1_yM}):
\begin{eqnarray}
\label{eq:omega}
\omega(y) = \tanh 2y + k.
\end{eqnarray}
Just like the $Z'$ contribution to $\sigma_2$, this is a sum of odd and even functions of $y$. Here $k$ is a numerical constant. We will adjust its value so that the contribution of one of the remaining three factors becomes negligible when integrated by $y$.

After the integration we obtain
\begin{eqnarray}
\label{eq:sigmastar}
\sigma^* - \sigma_\mathrm{SM}^* &=& \int_{-Y_m}^{Y_m} dy \, (\tanh 2y + k) \, \sigma_2 \nonumber \\
&=& \Bigg\{ \frac{\tanh y}{3} \left[ 12 a + b \, k \left(4 - \frac{1}{\cosh^2 y} \right) \right] \nonumber\\
&& \left. - 4 a \arctan (\tanh y) \Bigg\} \right|_{-Y_m}^{Y_m}, \nonumber\\
\sigma^* & = & \sigma_\mathrm{SM}^* + \bar{a}^2 \sigma_{\bar{a}^2}^* + \bar{a} \bar{v}_e \sigma_{\bar{a} \bar{v}_e}^* + \bar{a} \bar{v}_u \sigma_{\bar{a} \bar{v}_u}^*.
\end{eqnarray}
Note, that due to the symmetric integration region only the even part of the function $\omega(y)\sigma_2$ survives.
The factors $\sigma_{\mathrm{SM}}^*$, $\sigma_{\bar{a}^2}^*$, $\sigma_{\bar{a} \bar{v}_e}^*$, and $\sigma_{\bar{a} \bar{v}_u}^*$ are linear functions of $k$:
\begin{eqnarray}
\label{eq:sigmastar_bounds}
\sigma_{\mathrm{SM}}^* & = & (3.40 + 63.5 \, k) \,\, \mathrm{pb} \pm (0.39 + 5.4 \, k) \,\, \mathrm{pb}, \nonumber\\ 
\sigma_{\bar{a}^2}^* & = & (0.354 - 12.2 \, k) \,\, \mathrm{nb} \pm (0.003 - 1.2 \,k) \,\, \mathrm{nb}, \nonumber\\
\sigma_{\bar{a} \bar{v}_u}^* & = & (0.468 + 3.89 \, k) \,\, \mathrm{nb} \pm (0.009 + 0.17 \, k) \,\, \mathrm{nb}, \nonumber\\
\sigma_{\bar{a} \bar{v}_e}^* & = & (7.12 + 0.802 \, k) \, \mathrm{nb} \pm (0.52 + 0.068 \, k) \, \mathrm{nb}.
\end{eqnarray}

Let us construct an observable that is suitable for fitting of the axial-vector coupling $\bar{a}$ and the  coupling to the up-quark vector current, $\bar{v}_u$. That is, the factor $\sigma_{\bar{a} \bar{v}_e}^*$ has to be suppressed. We choose the suppression criteria
\begin{eqnarray}
\label{eq:k_criterium}
|\sigma_{\bar{a} \bar{v}_e}^*| < 0.01 |\sigma_{\bar{a}^2}^*|, \qquad |\sigma_{\bar{a} \bar{v}_e}^*| < 0.01 |\sigma_{\bar{a} \bar{v}_u}^*|
\end{eqnarray}
to calculate $k$ in Eq. (\ref{eq:sigmastar_bounds}). Overlap of the intervals obtained from the upper and lower bounds for factors gives the resulting interval $-9.18 \, \leq \, k \, \leq \, -8.55$. If we set $k$ = -9 in Eq. (\ref{eq:sigmastar_bounds}), the resulting observable will contain only two unknown $Z'$ parameters:
\begin{eqnarray}
\sigma^* & = & \sigma_\mathrm{SM}^* + \bar{a}^2 \sigma_{\bar{a}^2}^* + \bar{a} \bar{v}_u \sigma_{\bar{a} \bar{v}_u}^*, \nonumber\\
\sigma_{\mathrm{SM}}^* & = & -569 \pm 48 \,\, \mathrm{pb}, \nonumber\\ 
\sigma_{\bar{a}^2}^* & = & 111 \pm 10 \,\, \mathrm{nb}, \nonumber\\
\sigma_{\bar{a} \bar{v}_u}^* & = & -34.5 \pm 1.5 \,\, \mathrm{nb}.
\end{eqnarray}
This specific observable allows us to perform fitting of the $\bar{a}$ and $\bar{v}_u$ couplings.

There are two other possible observables in this approach: the one with suppressed $\sigma_{\bar{a} \bar{v}_u}^*$ and the one with suppressed $\sigma_{\bar{a}^2}^*$. However, the latter case cannot be realized in our scheme with suppression factor 0.01, because the intervals obtained for the lower and upper bounds from (\ref{eq:sigmastar_bounds}) do not overlap. Therefore, one has either to require weaker suppression in Eq. (\ref{eq:k_criterium}) or to narrow down the margin of error reducing the confidence level. Furthermore, this observable contains three $Z'$ couplings as opposed to two couplings in the case when $\sigma_{\bar{a}^2}$ or $\sigma_{\bar{a} \bar{v}_e}^*$ is suppressed. The mentioned flaws make this observable less attractive for data fitting, and we refrain from discussing it in the rest of our paper.

In Table \ref{tab:observables} we present the combinations of couplings that enter each of the proposed observables, together with the corresponding values of $k$ and $\sigma_{\bar{a}^2}^*$, $\sigma_{\bar{a} \bar{v}_u}^*$, and $\sigma_{\bar{a} \bar{v}_e}^*$. Note, that we choose certain $k$ values from the corresponding intervals.

\begin{table}[t]
\caption{\label{tab:observables}Couplings entering each of the two considered observables, together with the corresponding values of $k$, the SM contribution $\sigma_{\mathrm{SM}}^*$, and the factors $\sigma_{\bar{a}^2}^*$, $\sigma_{\bar{a} \bar{v}_u}^*$, and $\sigma_{\bar{a} \bar{v}_e}^*$.}
%\vskip3mm 
%\noindent
\begin{footnotesize}
%\centering
\begin{tabular}{lccccc}
\hline
\hline
couplings & $k$ & $\sigma_{\mathrm{SM}}^*$, pb & $\sigma_{\bar{a}^2}^*$, nb & $\sigma_{\bar{a} \bar{v}_u}^*$, nb & $\sigma_{\bar{a} \bar{v}_e}^*$, nb \\
\hline
$\bar{a}^2$, $\bar{a}\bar{v}_u$ & 
%(-9.32; -8.47) 
-9 & $-569 \pm 48$ & $111 \pm 10$ & $-34.5 \pm 1.5$ & suppressed \\
$\bar{a}^2$, $\bar{a}\bar{v}_e$ & 
%(-0.125; -0.116)
-0.12 & $-4.23 \pm 0.26$ & $1.82 \pm 0.14$ & suppressed & $7.02 \pm 0.52$ \\
\hline
\hline
\end{tabular}
\end{footnotesize}
\end{table}

The model-independent analysis of the LEP II data~\cite{GulovSkalozub:2009review,*GulovSkalozub:2010ijmpa} resulted in obtaining upper bounds for $\bar{a}^2$ and $\bar{v}_e^2$ at 95\% CL, both of order of $10^{-4}$. From Figs. \ref{fig:factors} (a), (b) and Table \ref{tab:observables} (see also \cite{GulovKozhushko:2011ijmpa}) it can be seen that these upper bounds are too large, since when substituted into Eq. \ref{eq:sigma2} they lead to a large deviation from the SM, which is not confirmed by any of the experimental data.
Therefore, we may expect at least some significant improvement of the LEP-motivated bounds.

Neither LEP data nor Tevatron or LHC data shows any explicit indications of the Abelian $Z'$. This provides motivation to investigate models with the so called leptophobic $Z'$ \cite{leptophobic:1,*leptophobic:2,*leptophobic:3,*leptophobic:4,*leptophobic:5,*leptophobic:6,*leptophobic:7,*leptophobic:8}. In these models $Z'$ boson couplings to the SM leptons are strongly suppressed compared to the quark couplings. Among other things, this parameterization allows to explain deviations of the precision electroweak data from the SM by introducing $Z'$ with the mass close to $m_Z$ \cite{Dermisek2011}. From the Lagrangian in Eq. (\ref{ZZplagr}) and the relations in Eq. (\ref{RGrel1}) it follows that in the leptophobic case $v_l$, $a_l$, and $a_q$ are small compared to $v_q$, and the leading $Z'$ contributions to the cross section are
\begin{eqnarray}
\label{eq:zpr_factors_leptophobic}
\sigma_\mathrm{DY} &=& \sigma_\mathrm{SM} + \sigma_{Z'}, \nonumber\\
\sigma_{Z'} &=& 
\bar{a} \bar{v}_u \sigma_{\bar{a} \bar{v}_u} + 
\bar{v}_u \bar{v}_e \sigma_{\bar{v}_u \bar{v}_e} \nonumber\\
&& + \bar{a} \bar{v}_c \sigma_{\bar{a} \bar{v}_c} + 
\bar{v}_c \bar{v}_e \sigma_{\bar{v}_c \bar{v}_e} + O(\bar{a}^2,\bar{a}\bar{v}_e).
\end{eqnarray}
After applying all the integrations discussed in Section \ref{sec:theobservable}, we end up with the observable where only the term $\bar{a} \bar{v}_u \sigma^*_{\bar{a} \bar{v}_u}$ survives. This observable is one-parametric: 
\begin{eqnarray}
\label{eq:sigmastar_leptophobic}
\sigma^* & = & \sigma_\mathrm{SM}^* + \bar{a} \bar{v}_u \sigma_{\bar{a} \bar{v}_u}^*.
\end{eqnarray}
The numerical values are the same as in the second line of Table \ref{tab:observables}.

Our results obtained for the dielectron case can be easily recalculated for dimuons, taking into account the difference between detector pseudorapidity coverages for electrons and muons. For example, the CDF Collaboration detects electrons with maximum pseudorapidity $|\eta_e| = 3.2$ \cite{CDF_DO_rapidity_ee} (the fiducial region is $|\eta_e| \leq 2.8$). For muons this value is $|\eta_{\mu}| = 1.5$ \cite{CDFrapidity_mumu}, therefore, the value of $Y_m$ for the $p\bar{p} \to \mu^+ \mu^-$ process is lower than for the dielectron case. This leads to different weight functions and $k$ values, which are presented in the Appendix\ref{app:numerical}.

\section{\label{sec:discussion}Discussion}

The data analysis performed by the LHC and Tevatron collaborations resulted in setting model-dependent lower bounds on the $Z'$ mass. In that analysis only the high-energy region of the Drell-Yan cross section was considered. In our paper we present a different approach that allows to search for a $Z'$ signal in the $p\bar{p} \to l^+ l^-$ process at the energies near $m_Z$.
In this region the most important contributions at the $Z$ peak come from the $Z-Z'$ mixing angle and $Z'$-induced contact couplings.
The $Z-Z'$ interference has to be taken into account, since it affects resonance shape as it was discussed in \cite{interference:1,*interference:2}.
The approach utilizes the model-independent relations between the effective $Z'$ couplings. Therefore, in case no signal is observed one would still be able to derive constraints for different $Z'$ models and compare them to the ones presented in \cite{CMS:2012zpr,*ATLAS:2011zpr}.

The proposed prescription includes the following steps:

1.~The triple-differential cross section of the Drell-Yan process is expressed in terms of three kinematic variables: the mass of an intermediate state, $M$, the intermediate-state rapidity $Y$, and the variable that describes the parton scattering subprocess, $y$. This cross section contains six unknown combinations of the $Z'$ couplings: $\bar{a}^2$, $\bar{a} \bar{v}_u$, $\bar{a} \bar{v}_l$, $\bar{a} \bar{v}_c$, $\bar{v}_u \bar{v}_l$, and $\bar{v}_c \bar{v}_l$;

2.~The cross section is integrated by $Y$ over the symmetric region $[-Y_m; Y_m]$ with the weight function $W(M,Y)$ defined in Eq.~(\ref{eq:observ_Y_wf}).
The integration limits have to be determined for each specific final state ($e^+ e^-$ or $\mu^+ \mu^-$) and detector individually. The function $A(M)$ has to be adjusted in such way, that the PDF factors for the second-generation quarks, $F_{c\bar{c}}(M)$ and $F_{s\bar{s}}(M)$, amount to less than 1\% of the PDF factor $F_{u\bar{u}}(M)$.
As a result we exclude $\bar{a} \bar{v}_c$ and $\bar{v}_c \bar{v}_l$ from the cross section;

3.~Integrate the cross section by $M$ over the $Z$ boson peak region: either $66 \mathrm{~GeV} \leq M \leq 116 \mathrm{~GeV}$ or $71 \mathrm{~GeV} \leq M \leq 111 \mathrm{~GeV}$, or any other region with bounds symmetric with respect to $m_Z$. These bounds have to be large enough, so that one could neglect the masses of the $u$, $d$, $c$, and $s$ quarks compared to $M$.
This integration suppresses the contribution of $\bar{v}_u \bar{v}_l$ to the cross section;

4.~The integration by $y$ with the properly adjusted weight function $\omega(y)$ from Eq.~(\ref{eq:omega}) allows to suppress either $\bar{a} \bar{v}_u$ or $\bar{a} \bar{v}_l$.

The obtained two alternative observables can be used in fitting the experimental data on the $p\bar{p} \to l^+ l^-$ scattering collected by the Tevatron collaborations. This allows to constrain the $Z'$ vector axial-vector couplings to SM fermions.

In case of the leptophobic $Z'$ boson, there is a one-parametric observable containing the combination of couplings $\bar{a}\bar{v}_u$.

There is a large amount of data on leptonic scattering processes collected in the LEP and LEP II experiments. The second observable in Table \ref{tab:observables} contains the coupling combinations $\bar{a}^2$ and $\bar{a} \bar{v}_e$ that also enter lepton scattering processes. It seems to be attractive for combined fits of the LEP and Tevatron data.

\appendix*

\section{\label{app:numerical}Numerical data for the $e^+ e^-$ and $\mu^+ \mu^-$ cases}
In this section we provide numerical values of the limits of integration by $Y$ and $y$, weight functions $W(M,Y)$ and $\omega(y)$, and the observable $\sigma^*$.

Generally, our approach is applicable for any final dileptonic state. However, the detector coverage is different for electrons and muons. The detector performance affects the proposed limits of integration by $M$, $Y$, and $y$. 
While the considered range for $M$ is well-covered by both CDF and D0 detectors, the detector-imposed limitations on the $Y$ and $y$ variables need a closer look. 

Both the D0 and CDF Collaborations detect electrons with maximum pseudorapidity $|\eta_e| = 3.2$ \cite{CDF_DO_rapidity_ee}. Therefore, the proposed integration limits $|Y| \leq 2.35$ and $|y|\leq 2.35$ are applicable in case of the $p\bar{p} \to e^+ e^-$ process. However, the pseudorapidity coverages for muons are $|\eta_\mu| \leq 2.0$ and $|\eta_\mu| \leq 1.5$ for D0 \cite{D0rapidity_mumu} and CDF \cite{CDFrapidity_mumu}, respectively. Because of this we have to appropriately adjust the limits of integration by $Y$ and $y$.

It is convenient to set $Y_1 = 0.75$ for all three cases, dielectrons at D0 and CDF, dimuons at D0, and dimuons at CDF. We choose the $Y_m$ value as shown in Table \ref{tab:Y_m}, since it represents the integration limit both for $Y$ and $y$. In Table \ref{tab:avsm} we present the values of $A(M)$, which enters the weight function $W(M,Y)$.

\begin{table}[h]
\caption{\label{tab:Y_m}Numerical values of $Y_m$ for different final-state dileptons at D0 and CDF.}
%\vskip3mm \noindent
\centering
\begin{tabular}{c@{\hskip 6mm}c@{\hskip 6mm}c@{\hskip 6mm}c}
\hline
\hline
~ & $e^+ e^-$ & \multicolumn{2}{c}{$\mu^+ \mu^-$} \\
\cline{3-4}
~ & ~ & D0 & CDF \\
\hline
$Y_m$ & 2.35 & 2.0 & 1.5 \\
\hline
\hline
\end{tabular}
\end{table}

\begin{table}[h]
\caption{\label{tab:avsm}Numerical values of $A(M)$ for different final-state dileptons at D0 and CDF.}
%\vskip3mm \noindent
\centering
\begin{tabular}{c@{\hskip 3mm}c@{\hskip 3mm}c@{\hskip 3mm}c}
\hline
\hline
$M$, GeV & $A(M)$, $e^+ e^-$ & \multicolumn{2}{c}{$A(M)$, $\mu^+ \mu^-$} \\
\cline{3-4}
~ & ~ & D$0$ & CDF \\
\hline
66 & -0.990053 & -0.929972 & -0.705828 \\
71 & -0.941569 & -0.884751 & -0.684274 \\
76 & -0.896807 & -0.844891 & -0.664090 \\
81 & -0.855355 & -0.809493 & -0.645150 \\
86 & -0.816857 & -0.777846 & -0.627341 \\
91 & -0.781010 & -0.749385 & -0.610566 \\
96 & -0.747549 & -0.723652 & -0.594736 \\
101 & -0.716242 & -0.700272 & -0.579775 \\
106 & -0.686890 & -0.678936 & -0.565613 \\
111 & -0.659313 & -0.659389 & -0.552186 \\
116 & -0.633356 & -0.641413 & -0.539441 \\
\hline
\hline
\end{tabular}
\end{table}

For interpolation we express $A(M)$ as
\begin{eqnarray}
A(M) = a_1 - \frac{a_2}{M + a_3}. \nonumber
\end{eqnarray}
The values of $a_1$, $a_2$, and $a_3$ are presented in Table \ref{tab:a1a2a3}.

\begin{table}[h]
\caption{\label{tab:a1a2a3}Numerical values of $a_1$, $a_2$, and $a_3$ for different final-state dileptons at D0 and CDF.}\vskip3mm \noindent
\centering
\begin{tabular}{c@{\hskip 3mm}r@{.}l@{\hskip 3mm}r@{.}l@{\hskip 3mm}r@{.}l}
\hline
\hline
~ & \multicolumn{2}{c}{$e^+ e^-$} & \multicolumn{4}{c}{$\mu^+ \mu^-$} \\
\cline{4-7}
~ & \multicolumn{2}{c}{~} & \multicolumn{2}{c}{D0} & \multicolumn{2}{c}{CDF} \\
\hline
$a_1$ & 0&225 & -0&212 & -0&049 \\
$a_2$ & 146& & 53&4 & 96&8 \\
$a_3$ & 54&3 & 8&35 & 81&3 \\
\hline
\hline
\end{tabular}
\end{table}

The weight function $w(y)$ from Eq. (\ref{eq:omega}), as well as the rest of the prescription, is used for both dimuonic cases. For the D0 case the resulting $k$-dependent factors are:
\begin{eqnarray}
\label{eq:sigmastar_bounds_D0}
\sigma_{\mathrm{SM}}^* & = & (2.84 + 50.8 \, k) \, \mathrm{pb} \pm (0.29 + 3.9 \, k) \, \mathrm{pb}, \nonumber\\ 
\sigma_{\bar{a}^2}^* & = & (0.251 - 9.86 \, k) \, \mathrm{nb} \pm (0.002 + 0.85 \,k) \, \mathrm{nb}, \nonumber\\
\sigma_{\bar{a} \bar{v}_u}^* & = & (0.349 + 3.00 \, k) \, \mathrm{nb} \pm (0.004 + 0.11 \, k) \, \mathrm{nb}, \nonumber\\
\sigma_{\bar{a} \bar{v}_\mu}^* & = & (5.54 + 0.648 \, k) \, \mathrm{nb} \pm (0.37 + 0.049 \, k) \, \mathrm{nb}. \nonumber
\end{eqnarray}
For the CDF case we have
\begin{eqnarray}
\label{eq:sigmastar_bounds_CDF}
\sigma_{\mathrm{SM}}^* & = & (1.33 + 23.3 \, k) \, \mathrm{pb} \pm (0.12 + 1.6 \, k) \, \mathrm{pb}, \nonumber\\ 
\sigma_{\bar{a}^2}^* & = & (0.090 - 4.57 \, k) \, \mathrm{nb} \pm (0.003 + 0.35 \,k) \, \mathrm{nb}, \nonumber\\
\sigma_{\bar{a} \bar{v}_u}^* & = & (0.139 + 1.31 \, k) \, \mathrm{nb} \pm (0.001 - 0.03 \, k) \, \mathrm{nb}, \nonumber\\
\sigma_{\bar{a} \bar{v}_\mu}^* & = & (2.36 + 0.302 \, k) \, \mathrm{nb} \pm (0.13 + 0.019 \, k) \, \mathrm{nb}. \nonumber
\end{eqnarray}

Values of $k$ obtained from suppression criteria (\ref{eq:k_criterium}) for D0 and CDF cases are shown in Table \ref{tab:observables_mumu}.
\begin{table}[h]
\caption{\label{tab:observables_mumu}Couplings entering each of the two considered observables, together with the corresponding values of $k$, the SM contribution $\sigma_{\mathrm{SM}}^*$, and the factors $\sigma_{\bar{a}^2}^*$, $\sigma_{\bar{a} \bar{v}_u}^*$, and $\sigma_{\bar{a} \bar{v}_e}^*$.}
%\vskip3mm \noindent
\centering
\begin{scriptsize}
\begin{tabular*}{1\linewidth}{lccccc}
\hline
\hline
couplings & $k$ & $\sigma_{\mathrm{SM}}^*$, pb & $\sigma_{\bar{a}^2}^*$, nb & $\sigma_{\bar{a} \bar{v}_u}^*$, nb & $\sigma_{\bar{a} \bar{v}_\mu}^*$, nb \\
\multicolumn{6}{c}{D0 case ($Y_m = 2.0$):}\\
\hline
$\bar{a}^2$, $\bar{a}\bar{v}_u$ & 
-8.5 & $-429 \pm 33$ & $84.0 \pm 7.2$ & $-25.2 \pm 0.9$ & suppressed \\
$\bar{a}^2$, $\bar{a}\bar{v}_\mu$ & 
-0.116 & $-3.06 \pm 0.17$ & ~$1.40 \pm 0.10$~~ & suppressed & $5.47 \pm 0.36$ \\
\hline
\multicolumn{6}{c}{CDF case ($Y_m = 1.5$):}\\
\hline
%couplings & $k$ & $\sigma_{\mathrm{SM}}^*$, pb & $\sigma_{\bar{a}^2}^*$, nb & $\sigma_{\bar{a} \bar{v}_u}^*$, nb & $\sigma_{\bar{a} \bar{v}_\mu}^*$, nb \\
%\hline
$\bar{a}^2$, $\bar{a}\bar{v}_u$ & 
-7.8 & $-180 \pm 12$ & $35.8 \pm 2.8$ & $-10.1 \pm 0.2$ & suppressed \\
$\bar{a}^2$, $\bar{a}\bar{v}_\mu$ & 
-0.106 & $-1.14 \pm 0.05$ & $0.575 \pm 0.034$ & suppressed & $2.33 \pm 0.13$ \\
\hline
\hline
\end{tabular*}
\end{scriptsize}
\end{table}

\bibliography{bibliography}

\end{document}